\documentclass[
 amsmath,amssymb,
 aps,showpacs,showkeys,twocolumn,
]{revtex4-1}
\usepackage{color}
\usepackage{graphicx}% Include figure files
\usepackage{dcolumn}% Align table columns on decimal point
\usepackage{bm}% bold math
\begin{document}

\preprint{APS/123-QED}

\title{Helical Edge Modes near Transition 
to Topological Insulator with Indirect Gap}% Force line breaks with \\

\author{Shijun Mao$^{1,2}$ }
% \altaffiliation[Also at ]{Department of Physics, Tsinghua
%University, Beijing 100084, P.R.China}%Lines break automatically or can be forced with \\
\author{Yoshio Kuramoto$^{1}$ }%
\affiliation{%
Department of Physics, Tohoku University,
Sendai 980-8578, Japan$^1$\\Department of Physics, Tsinghua
University, Beijing 100084, P.R.China$^2$
}%


 %\homepage{http://www.Second.institution.edu/~Charlie.Author}%affiliation{
%Department of Physics, Tsinghua
%University, Beijing 100084, P.R.China$^2$}%
%\affiliation{
%\affiliation{%


\date{\today}% It is always \today, today,

\begin{abstract}
Helical edge modes are characteristic of topological insulators 
in two dimensions. 
This paper demonstrates that helical edge modes remain across transitions to 
ordinary insulators or to semimetals under certain condition.
Straight and zigzag edges are considered in a tight-binding model on square lattice.
We focus on the case of indirect gap in bulk topological insulators,
and obtain the spectrum of edge modes 
on both sides of transitions.
For straight edge, the helical edge mode in topological insulators  
with strong particle-hole asymmetry has a reentrant region in momentum space. 
Edge modes show up even in ordinary insulators, but are absent in semimetals.
In zigzag edge,
the helical edge mode survives in both semimetals and ordinary insulators. 
However, the edge modes are absent inside the energy gap
of ordinary insulators.
All results are obtained analytically.

\end{abstract}

%Use showkeys class option if keyword
\maketitle
\section{Introduction}
In recent intensive study of topological insulators (TI), 
property of helical edge (2D TI) or surface (3D TI) states is 
one of the key topics.
It has been
shown that these states are robust against interaction and disorder \cite{ {he2}, {he3}}. 
The dissipationless spin current has potential application in
spintronics. 
In experiments for topological insulator materials, some materials are actually {\it bulk metals}, but still support topological surface states, such as Sb\cite{ex1,ex2},  Bi$_{0.91} $Sb$_{0.09}$\cite{ex3,ex4},\ Bi$_{2-x}$Mn$_{x}$Te$_3$\cite{ex5} and Bi$_2$Se$_3$ \cite{{ex6},{ex7}}. 
This kind of surface modes in (semi)metallic systems deserves further study,
especially as to their implication to the topologically nontrivial and trivial systems. 
With realistic energy bands for Bi$_2$Se$_3$, which are obtained by {\it ab initio} band calculation, 
the spectrum of helical surface modes has already been derived numerically \cite{zhang4}.  However, except for the low-energy part, 
further property of the surface mode, such as the existence region in momentum space, 
has not been discussed. 
Note that the topological argument about the robustness of the mode applies only to the energy region where bulk excitations are absent.
In this paper we are interested in global property of the helical modes including the region overlapping with bulk excitations.

One of the widely used models for 2D TI has been proposed by
Bernevig, Hughes and Zhang\cite{BHZ}, and is called the BHZ model. 
Helical edge states have been
further studied using the resultant continuum model \cite{{cm},
{cm2}} and the lattice regularized model \cite{{jpsj}, {imura},
{mao}}. 
In these previous papers, 
the particle-hole symmetry has been assumed.
Namely, the conduction and valence bands have the same spectrum except for the sign.
Hence the direct energy gap follows.
In materials such as Sb and Bi$_{0.91} $Sb$_{0.09}$, however, the energy gap or overlap involves different positions in momentum space.
Hence theoretical study beyond the particle-hole symmetric case is of practical interest.

In this paper, we study these helical states in both topologically nontrivial and trivial cases.
In order to clarify the characteristics associated with the particle-hole asymmetry in the 
simplest manner,
%full detail, 
we take the BHZ model on the square lattice, and derive the spectrum of helical modes analytically. 
We consider a
strip geometry of the 2D system with straight and zigzag edges as in
ref.\cite{mao}, and generalize its method for deriving helical edge states to particle-hole asymmetric case.

The paper is organized as follows: In \S 2, we review the particle-hole
asymmetric BHZ tight-binding model, paying attention to both direct
and indirect gap-closing in 2D Brillouin zone (BZ). 
Sections 3 and 4 are devoted to 
the straight edge and zigzag edge systems, respectively. We
analytically derive spectrum of helical edge states in particle-hole asymmetric
system. 
In straight edge case, edge mode is present in TI and, under appropriate conditions, also in ordinary insulators (OI). However, it's absent in semimetals (SM). 
In zigzag edge case, helical edge states survive in each case of TI, OI, and SM. 
The summary and outlook are given in \S 5.

\section{Particle-hole asymmetric BHZ model}
We consider the BHZ model given by the following block diagonalized $4\times 4$ matrix:
\begin{align}
&H (\vec k)=\left[
\begin{array}{cc}
h(\vec k)  &   0 \\
0  &   h^*(-\vec k) \\
\end{array}
\right], \label{Htot}
\end{align}
where $\vec k=(k_x,k_y)$ is a 2D crystal momentum.
%, measured from $\Gamma$-point. 
The lower-right
block $h^* (-\vec k)$ for down-spin 
%is a $2\times 2$ matrix, and
is deduced
from the upper-left block
$h(\vec k)$ for up-spin by time-reversal transformation.
We parametrize 
$h(\vec k)$ as
\begin{equation}
h (\vec k) = \epsilon(\vec k) \sigma_0+{\vec d} (\vec k) \cdot {\vec \sigma} = \left[
\begin{array}{cc}
\epsilon+d_z  &  d_x -i d_y  \\
d_x  +i d_y &  \epsilon- d_z
\end{array}
\right], \label{hk}
\end{equation}
where $\sigma_0$ is the $2\times 2$ unit matrix in the pseudo-spin space representing the two kinds of orbitals,
and $\vec\sigma$ is the vector composed of the Pauli matrices $\sigma_x, \sigma_y$ and $\sigma_z$.  We write $\sigma_0=1$ hereafter unless confusion arises.
%\begin{align}
%\sigma_x &=\left(\begin{array}{cc}
%0  &  1  \\
%1 &  0
%\end{array}\right), \quad
%\sigma_y=\left(\begin{array}{cc}
%0  &  -i  \\
%i &  0
%\end{array}\right),\nonumber \\
%\sigma_z &=\left(\begin{array}{cc}
%1  &  0  \\
%0 &  -1
%\end{array}\right).
%%\sigma_0=\left(\begin{array}{cc}
%%1  &  0  \\
%%0 &  1
%%\end{array}\right),
%\end{align}
We consider the BHZ model over the whole BZ of the
square lattice. Then 
the energy parameters are given by the tight-binding form as
\begin{align}
\epsilon({\vec k}) = C-2D(2-\cos k_x - \cos k_y) 
\label{epsilon-k}\\
d_x ({\vec k})  = A\sin k_x, \quad
d_y ({\vec k})  = A\sin k_y, \\
d_z ({\vec k})  = \Delta -2B (2-\cos k_x-\cos k_y),
\end{align}
with the lattice constant $a$ set to unity\cite{BHZ}.

The bulk energy $E_{\rm b}({\vec k})$
is obtained as
\begin{align}
E_{\rm b\pm}({\vec k})
= &\epsilon({\vec k})\pm  \left\{
A^2\left( \sin^2 k_x+\sin^2 k_y \right)\right. \nonumber \\
+&\left.
\left[ \Delta-4B+2B\left(
\cos k_x+\cos k_y \right) \right]^2 \right\}^{1/2},
 \label{bulk}
\end{align}
It is clear that finite $\epsilon({\vec k})$ breaks the particle-hole symmetry.   Since $C$ in Eq.(\ref{epsilon-k}) merely describes the energy shift of the origin, we put $C=0$ for simplicity.  
%The parameter $D$ controls different shapes of conduction and valence bands.
As in our previous paper\cite{mao}, we use $B$ as the unit of energy and put $B=1$ hereafter.

The change of $\Delta$ will cause
transition between TI and OI.
With inversion symmetry, the direct gap can close only at time-reversal invariant points, namely at $\Gamma=(0,0)$, $X=(\pi,0)$, $X'=(0,\pi)$ and
$M=(\pi,\pi)$ \cite{gapless}, where bulk energies become
\begin{align}
E_{\rm b\pm}(0,0)&=\pm |\Delta|, \label{00}\\
E_{\rm b\pm}(0,\pi)&=-4D \pm |\Delta-4|=E_{\rm b\pm}(\pi,0), \label{10}\\
E_{\rm b\pm}(\pi,\pi)&=-8D \pm |\Delta-8|. \label{11}
\end{align}
As one varies the 
%mass 
parameter $\Delta$,
the direct gap closes at
$\Gamma$, for example, when $\Delta=0$.
Similarly the gap closes at $X$ (also $X'$) with $\Delta=4$,
and at $M$ with $\Delta=8$.

Due to finite $\epsilon(\vec k)$, it is also possible to have
the indirect gap-closing.
For example, we obtain 
$E_{\rm b+}(0,\pi)=-\Delta$ with $0<\Delta<4$ and $D=1$ from Eq.(\ref{10}), 
which becomes the same as 
$E_{\rm b-}(0,0)=-\Delta$.
Then the transition occurs between TI and SM at $D=1$ for 
$0<\Delta<4$.

%Without loss of generality, 
In this paper we only deal with edge mode with up-spin
and confine to the case $A,D>0$ and $\Delta<4$,
since the down-spin part is obtained by using the time-reversal symmetry.
Note that the result for negative $D$ is obtained by changing the role of conduction and valence bands.  
On the other hand, the case $\Delta>4$ is understood by changing the role of $\Gamma$ and $M$ points in the Brillouin zone\cite{mao}.

\section{Straight edge}
%%%%%%%%%%%%%%%%%%%%%%%%%%%%%%%%%%%%%%%%%%%%%%%%%%%%%%%%%%%%%%%
\begin{figure}%[ht]
\begin{center}
\includegraphics[width=0.4\linewidth]{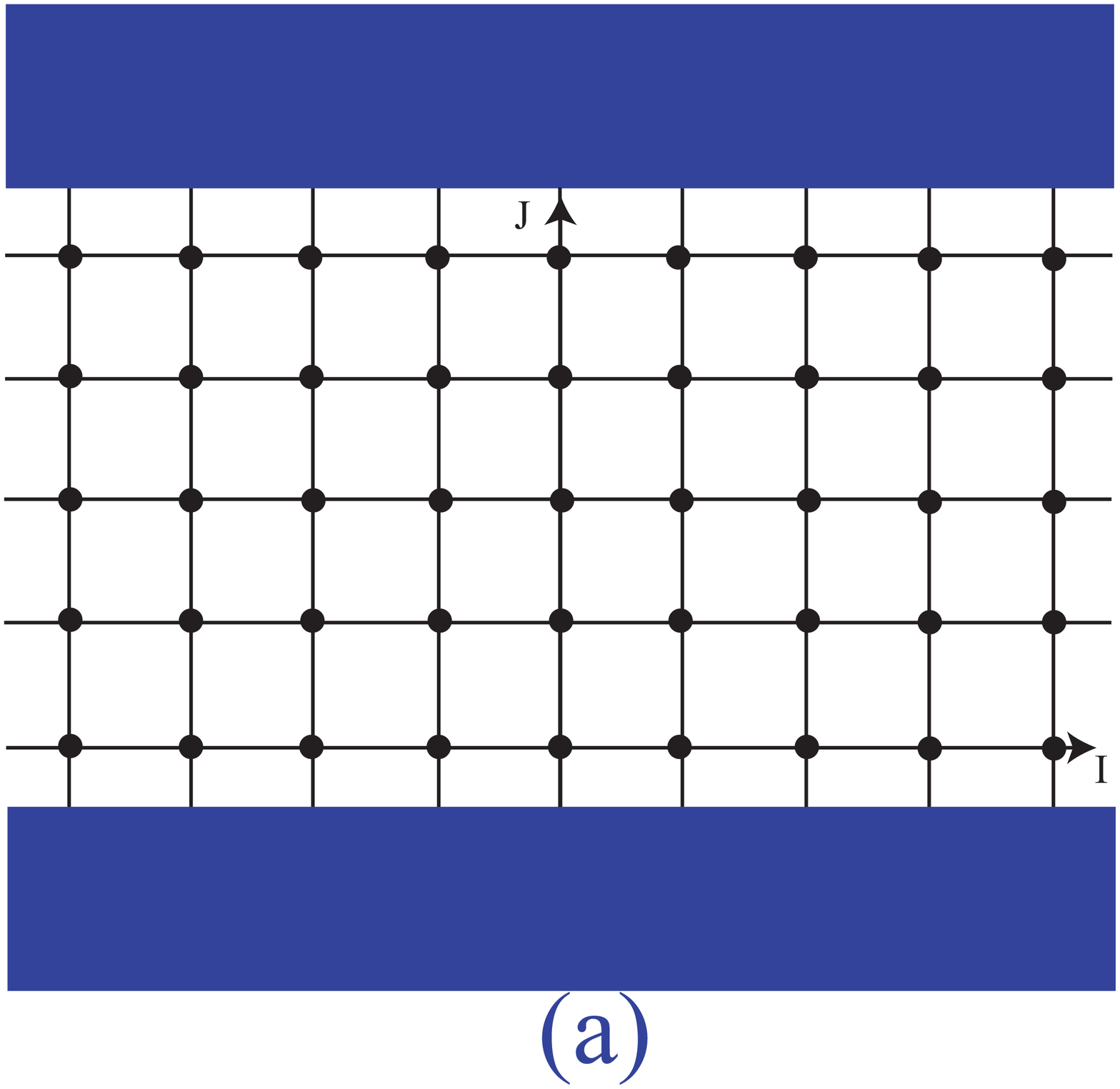}
\includegraphics[width=0.415\linewidth]{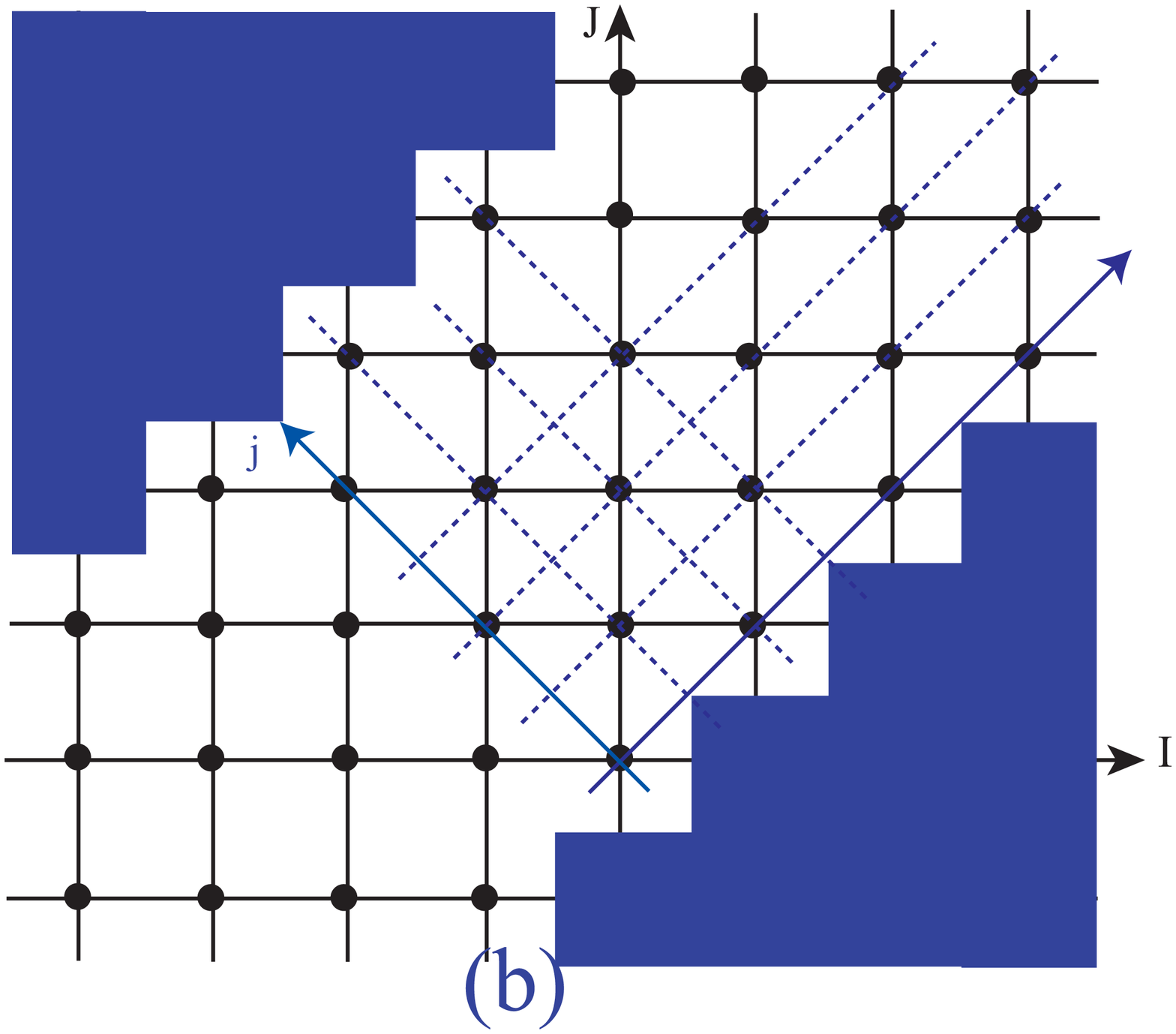}
\caption{(Color online) (a) Straight edge system with boundaries in (1,0) direction. (b)Zigzag edge system with boundaries in
(1,1) direction.} \label{fig1}
\end{center}
\end{figure}
%%%%%%%%%%%%%%%%%%%%%%%%%%%%%%%%%%%%%%%%%%%%%%%%%%%%%%%%%%%%%%%%

\subsection{Analytic solution of spectrum and wave function}
Let us first consider a straight edge geometry (see Fig. \ref{fig1}a), in which
electrons are confined to $N_{r}$ rows in a strip between $y=1$ and
$y=N_{r}$, i.e., the two edges are along $x$-axis. Translational
invariance along $x$-axis allows for constructing a 1D tight-binding
Hamiltonian 
$h_{01}(k)$ with $k=k_x$
in terms of the 
%two-component 
creation $c^\dagger_{J}(k)$ 
and annihilation $c_{J}(k)$ operators 
with row index $J$, which are
two-component vectors in the pseudo-spin space.
For up spin, the 1D model $h_{\uparrow}$
is obtained as
\begin{align}
h_{\uparrow} &=\sum_{k} h_{01}(k), \\
h_{01}(k) &= \sum_{J} c^\dagger_J(k) \hat{\cal E} (k) c_J(k)
\nonumber \\& + \sum_{J} \left[ c^\dagger_{J}(k) \hat{t}_y
c_{J+1}(k) + h.c. \right] \label{hNR}
\end{align}
where $\hat{\cal E} (k)$ and $\hat{t}_{y}$ are given by
\begin{align}
\hat{\cal E} (k) =&\left(-4D+2D \cos k\right) \nonumber \\
&+A \sin k \sigma_x + \left( \Delta_B +2 \cos k \right)
\sigma_z,\\
\hat{t}_{y}=&-i\frac{A}{2}\sigma_y+ \sigma_z+D. \label{epsilon}
\end{align}
Here and in the following,
we use the notation $\Delta_B \equiv \Delta-4 $.
%Then, eq.(\ref{bulk}) is written as
%\begin{align}
%E_{\rm b}({\vec k})^2 &= A^2\left( \sin^2 k_x+\sin^2 k_y \right)\nonumber \\  &+\left[ \Delta_4+2\left(
%\cos k_x+\cos k_y \right) \right]^2
%\end{align}
The corresponding Schr\"{o}dinger equation is given by
\begin{equation}
\hat{\cal E}(k)\Psi_J +\hat t_y \Psi_{J-1}+\hat t _y^\dagger
\Psi_{J+1} = E (k)\Psi_J \label{TB2}
\end{equation}
where $\Psi_J$ is the two-component amplitude with row index $J$.
The straight edges along the $J=1$ row and $J=N_{r}$ row can be
implemented by open boundary condition $\Psi_0=\Psi_{N_r+1}=0$.

We consider the thermodynamic limit $N_r\rightarrow \infty$, and 
try an edge state solution with property: $\Psi_{J+1} =
\lambda\Psi_J=\lambda^{J+1}\Psi$ with
$|\lambda|<1$\cite{{Ed1},{Ed2}}. 
Here $\lambda$ is a complex number in general. 
Then Eq.(\ref{TB2}) can be
written in the following form:
\begin{equation}
\left[  \hat{\cal E}(k) +\lambda \hat t _y^\dagger +\lambda^{-1}
\hat t_y \right] \Psi \equiv P_{01}(\lambda, k)\Psi =E(k)\Psi.
\label{TB3}
\end{equation}
%We can conclude
%The Hermitian part
%only comes from $\hat{\cal E}(k)$.
%Therefore, we 
%Since $P_{01}(\lambda,k)$ is non-Hermitian, 
Although $P_{01}(\lambda,k)$ is a $2\times 2$ matrix,
there is at most one solution $\Psi$ for the helical edge mode in Eq.(\ref{TB3}).
We decompose 
$P_{01}(\lambda,k)$ into the Hermitian part and the rest called the annihilator\cite{mao}.
The absence of the particle-hole symmetry in the present case requires
a little more complicated procedure than before \cite{mao}.
Let us first introduce a parameter $\phi$ and make the following
transformation:
\begin{align}
&\sigma_{X}=\cos{\phi} \, \sigma_{x}+ \sin{\phi}\,\sigma_{z}\\
&\sigma_{Z}=-\sin{\phi}\,\sigma_{x}+ \cos{\phi}\,\sigma_{z}
%\\&\sigma_{2}=\sigma_{y}
\end{align}
Then, we have
\begin{align}
&\hat{\cal E}(k)={\cal E}_0  + {\cal E}_1 \sigma_X +{\cal E}_3 \sigma_Z,\\
&{\cal E}_0=-4D+2D \cos{k},\\
&{\cal E}_1=A \sin{k} \cos{\phi}+ \left(\Delta_B+2  \cos{k}\right) \sin{\phi},\\
&{\cal E}_3=-A \sin{k} \sin{\phi}+\left(\Delta_B+2  \cos{k}\right) \cos{\phi},
\label{E3}\\
&
\lambda {\hat t}^{\dagger}_{y} + \lambda^{-1} {\hat t}_{y} %\nonumber \\ 
=\frac{i}{2}A\left(\lambda-\lambda^{-1}\right) \sigma_y+  \cos{\phi}\left(\lambda+\lambda^{-1}\right) \sigma_Z\nonumber
\\
&
+\left(\lambda+\lambda^{-1}\right)\left( \sin{\phi}\,\sigma_X+ D
\right).
\end{align}
Now we decompose $P_{01}(\lambda,k)$ in Eq.(\ref{TB3}) as 
\begin{align}
P_{01}=H_{01}+F_{01},
\end{align}
where 
the Hermitian part $H_{01}$ is chosen as combination of $\sigma_0\ (=1)$
and $\sigma_{X}$ terms. 
The eigenstate $\Psi$ of $H_{01}$ obeys the relation $\sigma_X
\Psi=s \Psi$ with $s=\pm 1$.
Furthermore, $\sigma_y$ and $\sigma_Z$ terms combine to form the 
annihilator $F_{01}$ \cite{mao}.  Namely we obtain
\begin{align}
&H_{01}\Psi =E(k) \Psi, \quad F_{01}\Psi=0,\\
&H_{01}={\cal E}_0 +{\cal E}_1 \sigma_X+\left(\lambda+\lambda^{-1}\right)\left( \sin{\phi}\,\sigma_X+ D \right),\\
&F_{01}=i\frac{A}{2}\left(\lambda-\lambda^{-1}\right) \sigma_y+\left[{\cal E}_3+  \cos{\phi}\left(\lambda+\lambda^{-1}\right)\right] \sigma_Z,
\label{F01}
\end{align}

In view of $E(k)$ being real, and the parameter $\lambda$ being complex in general,
we require in $H_{01}$ 
the condition
\begin{align}
\left( \sin{\phi}\,\sigma_X+ D \right) \Psi=0,
\end{align}
which determines $\phi$ as
\begin{align}
\sin{\phi}=-s{D}.
\label{theta}
%&\cos{\phi}=\sqrt{1-D^2}
\end{align}
Then $E(k)$ of the edge mode reads
\begin{align}
E(k)=s{\cal E}_1 +{\cal E}_0.
%,s=\pm1
\end{align}
The condition for $F_{01}$, which is given by Eq.(\ref{F01}),
to form the annihilator $\sigma_y+is \sigma_Z$ reads
\begin{align}
{\cal E}_3+ \cos{\phi}\left(\lambda+\lambda^{-1}\right)=-s\frac{A}{2}\left(\lambda-\lambda^{-1}\right),
\end{align}
which determines $\lambda$ since $\phi$ has already been fixed by Eq.(\ref{theta}).
The solutions $\lambda=\lambda_{s\pm}(k)$
are given by
\begin{align}
\lambda_{s\pm}(k)=\frac{-{\cal E}_3 \pm \sqrt{{\cal E}^2_3-4\cos^2{\phi}+A^2}}{2 \cos{\phi}+s A},
\label{lambda10}
\end{align}
with $\cos{\phi}=\sqrt{1-D^2}$.
The dependence on $k$ comes only through ${\cal E}_3$.
For complex $\lambda$, we obtain
\begin{align}
|\lambda_{s\pm}|^{2}=\frac{2  \sqrt{1-D^2}-s A}{2\sqrt{1-D^2}+s A},
\end{align}
which does not depend on $k$, and  
should be less than unity.
Therefore, we must choose $s=1$ that corresponds to
the right-going mode. 
We obtain the spectrum:
\begin{align}
E_{\uparrow}(k)={\cal E}_1 +{\cal E}_0
= -D\Delta+A\sqrt{1-D^2}\sin k.
\end{align}
In the case of $D=0$, the spectrum is reduced to $A\sin k$, which was already obtained 
\cite{jpsj,imura,mao}.
The constant term $-D\Delta$ can be canceled by the chemical potential.
Then the zero mode occurs at $k=0$.
Otherwise the zero mode has a finite momentum.
The two-component wave function is given by
\begin{align}
\Psi_{J}=\left( 
\lambda^{J}_{+}- \lambda^{J}_{-} \right)\left(\begin{array}{cc}
\sqrt{1-D} \\
\sqrt{1+D}
\end{array}
\right),
\end{align}
apart from the normalization factor. 
Hereafter we use notation $\lambda_{\pm}=\lambda_{1 \pm}$, since only the case of $s=1$ is relevant.

\begin{figure}%[ht]
\begin{center}
\includegraphics[width=0.8\linewidth]{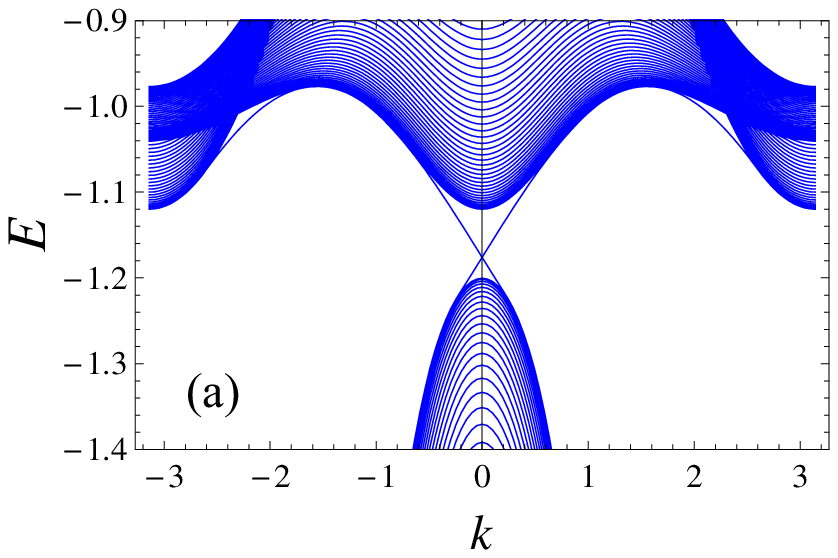}\\
\includegraphics[width=0.8\linewidth]{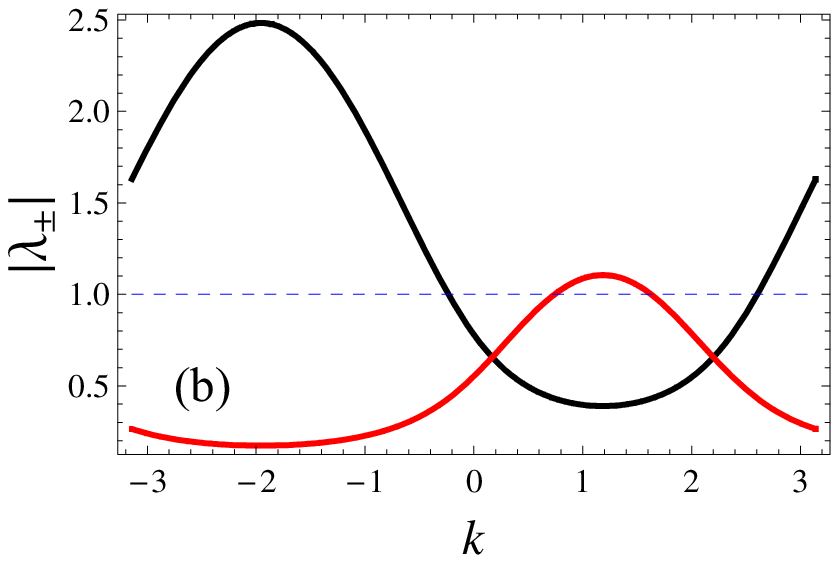}
\caption{ (Color online) (a) Energy band $E(k)$ and (b) parameters 
$|\lambda_{-}|$ (red) and
$|\lambda_{+}|$ (black) in
straight edge system with $\Delta=1.2$, $A=1$, $D=0.98$.
Here, parameter $|\lambda_{\pm}|$ corresponds to right-going branch
of edge mode. } \label{fig2}
\end{center}
\end{figure}

Fig.\ref{fig2} (a) shows the spectrum of the system including both 
$h_{\uparrow}$ and $h_{\downarrow}$, {\it i.e.}, two helical modes.
The bulk spectrum is obtained by projecting the 2D spectrum for a finite but large system into 1D BZ of the straight edge.
The parameter $|\lambda_{\pm}|$ is shown in Fig.\ref{fig2} (b).
The edge modes are present only if $|\lambda| < 1$ is satisfied for both $\lambda_{\pm}$.
It is clear from Fig.\ref{fig2}(b) that the 
momentum range for the edge mode is split into two;
one is localized around center of BZ, and the other allows for a reentrant edge
mode \cite{{imura}, {mao}}. 
%%%%%%%%%%%%%%%%%%%%%%%%%%%%%%%%%%%%%%%%%%%%%%%%%%%%%%%%%%%%%%%
%%%%%%%%%%%%%%%%%%%%%%%%%%%%%%%%%%%%%%%%%%%%%%%%%%%%%%%%%%%%%%%%
Furthermore,  
the two parts have exactly the same width for the momentum range. 
%In the region with $|\lambda_{\pm}|\geq 1$, edge mode degenerates with lowest bulk energy band.
This fact can be understood by writing ${\cal E}_3$ in Eq.(\ref{E3}) as
\begin{align}
&{\cal E}_3 -  \Delta_B\sqrt{1-D^2} = 
AD\sin k +2\sqrt{1-D^2}\cos k \nonumber\\
&=\sqrt{A^2D^2+4(1-D^2)}\sin (k+\alpha),
\label{E_3}
\end{align}
with $\tan\alpha = 2\sqrt{1-D^2}/(AD)$.   
This expression shows that 
${\cal E}_3$ is symmetric about 
\begin{align}
k=k_c = \pi/2-\alpha = \arctan \frac{AD}{2\sqrt{1-D^2}},
\label{kc}
\end{align}
which becomes $1.18$ with the chosen values of $A=1$ and $D=0.98$.
Hence, according to Eq.(\ref{lambda10}), the value of $\lambda_{\pm}$ is also symmetric around $k_c$ as seen from Fig.\ref{fig2}(b).
The allowed momentum range for the edge mode with $|\lambda_{\pm}|<1$
is also symmetric around $k_c$.

\subsection{Transition to semimetals}

\begin{figure}%[ht]
\begin{center}
\includegraphics[width=0.8\linewidth]{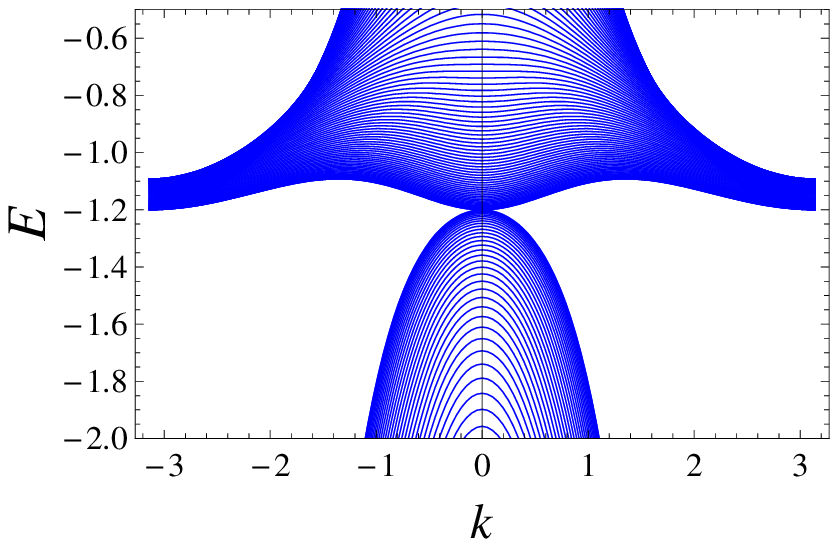}\\
\includegraphics[width=0.8\linewidth]{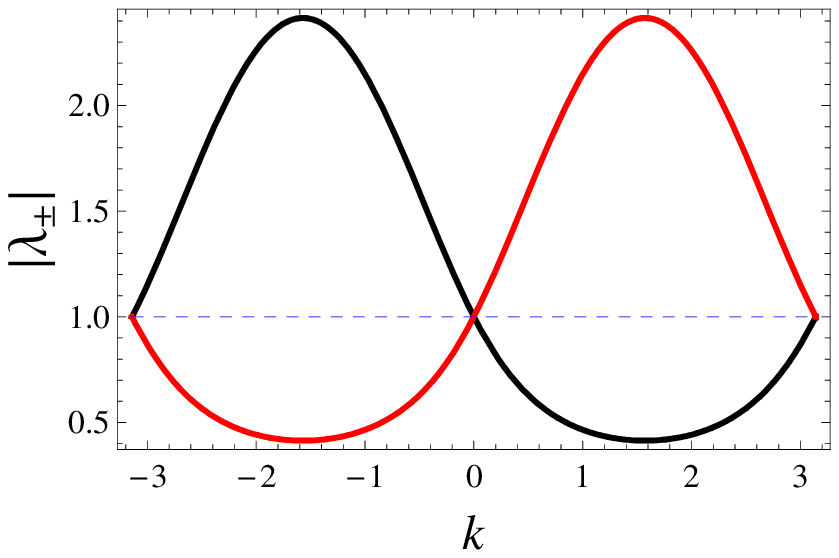}
\caption{ (Color online) Energy band $E(k)$ and parameter $|\lambda_{\pm}|$ in
straight edge system with $\Delta=1.2$, $A=1$, $D=1.0$.
Here, parameter $|\lambda_{\pm}|$ corresponds to right-going branch
of edge mode. } \label{fig3}
\end{center}
\end{figure}

Fig.\ref{fig3} shows the critical case with $D=1$
where the minimum value of conduction band is equal to the maximum value of valence band:
\begin{align}
&E_{\rm b+}|_{min}=E_{\rm b+}(0,\pi)=E_{\rm b+}(\pi,\pi)=-\Delta\\
&E_{\rm b-}|_{max}=E_{\rm b-}(0,0)=-\Delta.
\end{align}
For the parameter $\lambda_{\pm}$, we obtain
\begin{align}
&\lambda_{\pm}=-\sin k \pm \sqrt{1+\sin ^2 k},
\label{lambda10c}
\end{align}
which means 
%\begin{align}&
$\lambda_{-}\leq-1$ for  $0 \leq k \leq \pi$, and 
$\lambda_{+}\geq 1$ for  $-\pi \leq k \leq 0$. 
Therefore, the edge mode is absent for any momentum
in the critical case.

%%%%%%%%%%%%%%%%%%%%%%%%%%%%%%%%%%%%%%%%%%%%%%%%%%%%%%%%%%%%%%%
%%%%%%%%%%%%%%%%%%%%%%%%%%%%%%%%%%%%%%%%%%%%%%%%%%%%%%%%%%%%%%%%

\begin{figure}%[ht]
\begin{center}
\includegraphics[width=0.8\linewidth]{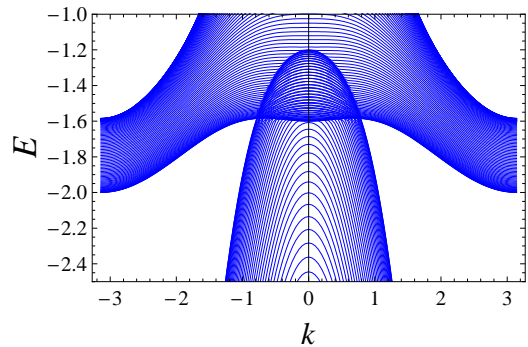}
\caption{ (Color online) Energy band $E(k)$ of
straight edge system with $\Delta=1.2$, $A=1$, $D=1.1$.} 
\label{fig4}
\end{center}
\end{figure}

For $D>1$,  the bulk system becomes a semimetal. 
Fig.\ref{fig4} shows the spectrum in this case with $D=1.1$.
Because we have $\sin{\phi}=-D <-1$, the edge mode is absent.

Let us summarize the situation 
with changing $D$ for the straight edge: 
With $D=0$, the system is topologically nontrivial
insulator with particle-hole symmetry. 
For $D>0$, the particle-hole
symmetry is broken and the shape of valence and conduction bands become different.
As getting close to the critical case with $D=1$, the spectrum of the edge mode
separates into two parts with the same width in the 1D momentum space, and both
parts shrink simultaneously with further increase of $D$. 
At $D=1$, associated with closing of the bulk band gap,
edge modes disappear in the whole BZ. 
With $D>1$, the system enters into semimetals where no edge modes are present for the straight edge.

%%%%%%%%%%%%%%%%%%%%%%%%%%%%%%%%%%%%%%%%%%%%%%%%%%%%%%%%%%%%%%%
%%%%%%%%%%%%%%%%%%%%%%%%%%%%%%%%%%%%%%%%%%%%%%%%%%%%%%%%%%%%%%%%

\subsection{Edge mode in ordinary insulators}

We consider the edge mode in ordinary (topologically trivial) insulators (OI).
In the particle-hole symmetric system, there is no edge modes \cite{{BHZ},{jpsj}}. 
In the present case,
edge modes can show up even in OI depending on the value of $D$.
The presence of edge modes in OI can be analyzed conveniently by considering
two limiting cases:
(i) particle-hole symmetric
case with $D=0$, where we obtain  
${\cal E}_{3}=\Delta_B+2  \cos {k}$;
% and $|\lambda_{+}|>|\lambda_{-}|$; 
(ii) critical case with $D=1$, where the bulk energy
gap closes and edge mode disappears.  Here we obtain ${\cal E}_{3}=A
\sin {k}$.
In order to derive the momentum region allowing for edge modes,
we consider extrema of $\lambda_{\pm}$ at $k=k_c$.
We obtain from Eq.(\ref{E_3})
\begin{align}
{\cal E}_3 =  \Delta_B\sqrt{1-D^2} +\sqrt{A^2D^2+4(1-D^2)},
\end{align}
which is negative for 
\begin{align}
\Delta_B <-\left( 
\frac{A^2D^2}{1-D^2}+4 \right)^{1/2}.
\end{align}
%The parameters in the following calculation satisfies this condition, except for a case of $D\sim 1$ as in Figs.\ref{fig2} and \ref{fig5b}.
In the case of ${\cal E}_3 <0$ we obtain $|\lambda_{+}(k_c)| > |\lambda_{-}(k_c)|$,
and otherwise opposite inequality.

\begin{figure}%[ht]
\begin{center}
\includegraphics[width=0.8\linewidth]{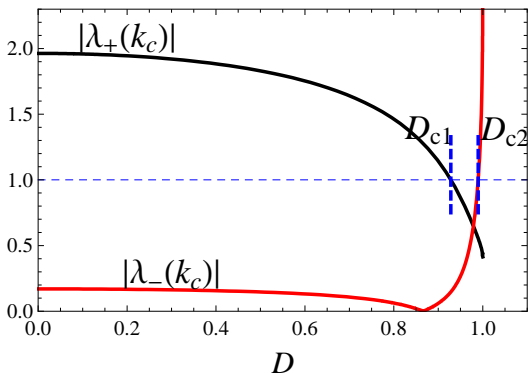}
\caption{ (Color online) $|\lambda_{\pm}(k_c)|$ as a function of $D$, with $\Delta=-1.2$ and $A=1$.  Helical edge modes appear for $D_{c1} <D< 1$
although system is an ordinary insulator. Especially, for $D_{c2} <D <1$, helical edge mode contains two separate parts.
We obtain 
$D_{c1}=0.93$ and
$D_{c2} =0.99$
with present choice of parameters.}
\label{fig6}
\end{center}
\end{figure}

Fig.\ref{fig6} plots the parameters $|\lambda_{\pm}(k_c)|$ given by Eq.(\ref{lambda10})
as a function of $D$.  
We set 
$\Delta=-1.2$, or
$\Delta_B=-5.2$,
which makes the system OI.  
Helical edge modes appear for 
$D_{c1}<D<1$ where
%$D_{c1}<D<D_{c2}$ where $|\lambda_{\pm}|<1$ for both + and $-$.  
the boundary $D_{c1}$ are given by solution of the condition
\begin{align}
\lambda_{+}(k_c)=1.
\end{align}
With $D$ increasing from zero to unity,
$|\lambda_{+}(k_c)|$ decreases, as seen in Fig {\ref{fig6}}.
As long as $D<D_{c1}$, we have the relation $|\lambda_{+}(k_c)|>1$,
and no edge modes.

%%%%%%%%%%%%%%%%%%%%%%%%%%%%%%%%%%%%%%%%%%%%%%%%%%%%%%%%%%%%%%%
%%%%%%%%%%%%%%%%%%%%%%%%%%%%%%%%%%%%%%%%%%%%%%%%%%%%%%%%%%%%%%%%

\begin{figure}%[ht]
\begin{center}
\includegraphics[width=0.8\linewidth]{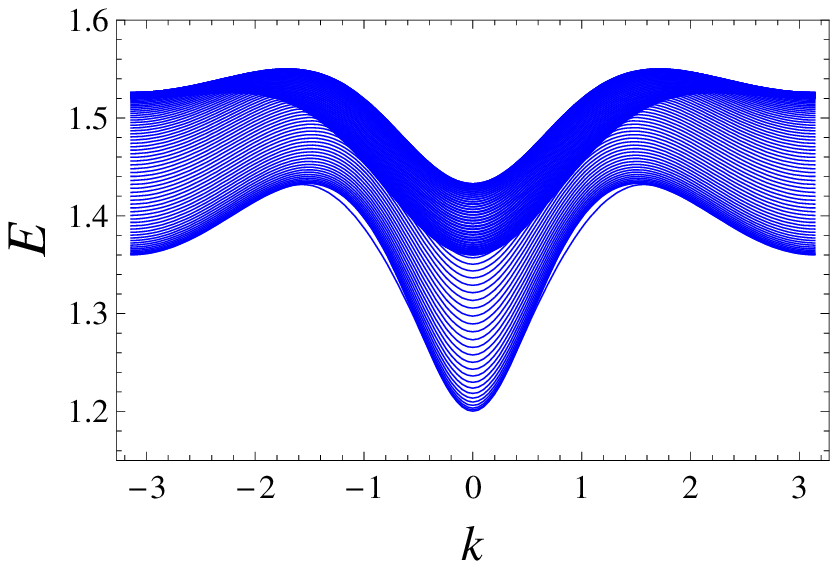}
\includegraphics[width=0.8\linewidth]{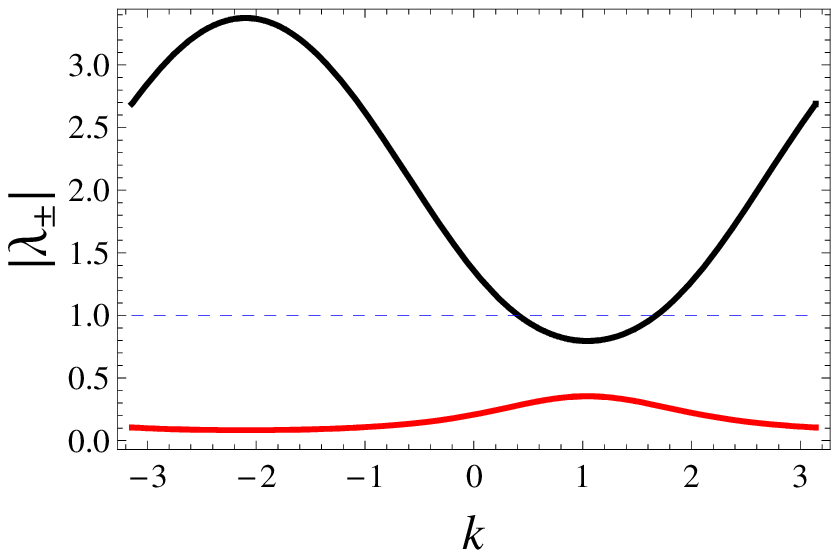}
\caption{ (Color online) Conduction band $E(k)$ and parameter $|\lambda_{\pm}|$ for $D=0.96$, $\Delta=-1.2$, $A=1$.  
The parameter $|\lambda_{\pm}|$
corresponds to right-going branch of edge mode,
which does not have the reentrant behavior. } 
\label{fig5}
\end{center}
\end{figure}

At $D=D_{c1}$, we obtain 
$|\lambda_{-}(k_c)|<|\lambda_{+}(k_c)|=1$, and the edge mode begins to show
up. 
Fig.{\ref{fig5}} shows an example of the spectrum in the OI
with $\Delta=-1.2$ and $A=1$. 
Here edge modes appear in the middle of the BZ without the reentrant behavior.  
%%%%%%%%%%%%%%%%%%%%%%%%%%%%%%%%%%%%%%%%%%%%%%%%%%%%%%%%%%%%%%%
%%%%%%%%%%%%%%%%%%%%%%%%%%%%%%%%%%%%%%%%%%%%%%%%%%%%%%%%%%%%%%%%
\begin{figure}%[ht]
\begin{center}
\includegraphics[width=0.8\linewidth]{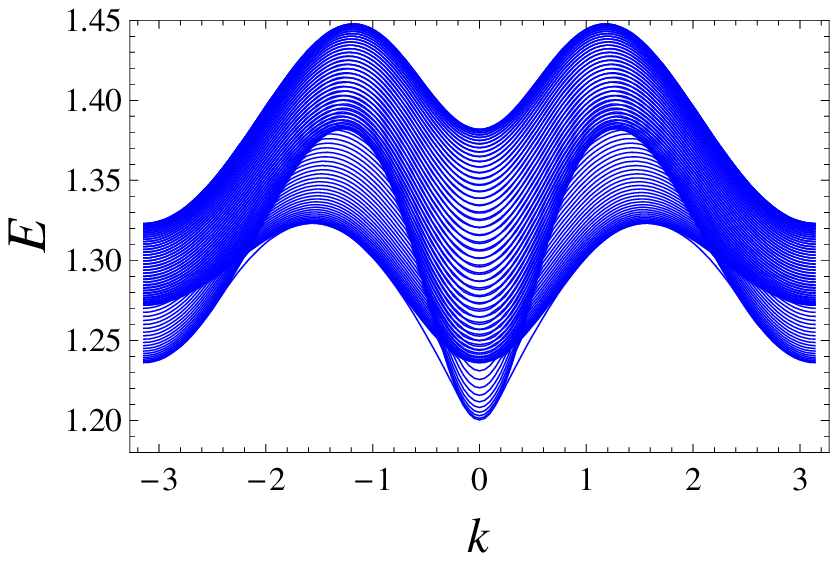}
\includegraphics[width=0.8\linewidth]{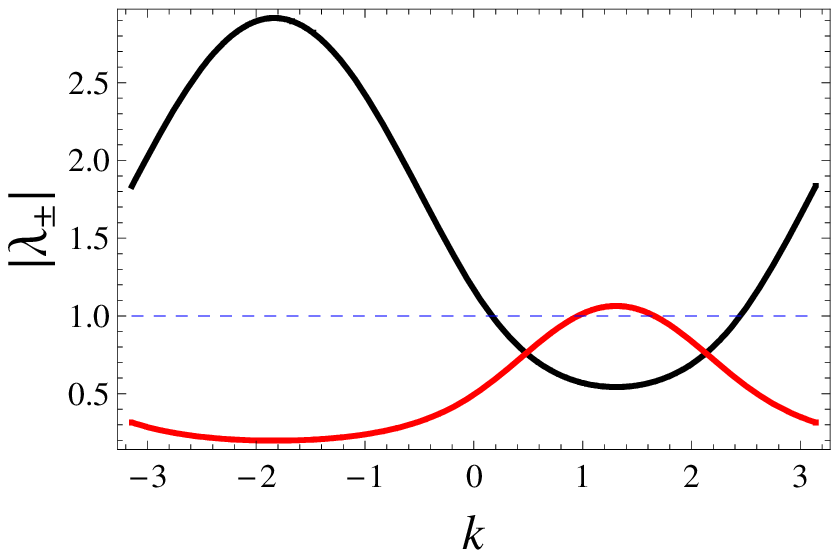}
\caption{ (Color online) Conduction band $E(k)$ and parameter $|\lambda_{\pm}|$ for 
%$D=0.96$ (upper two panels) and 
$D=0.991$.   Other parameters are the same as those in Fig.\ref{fig5}.
Reentrant modes are present in this case.
}
\label{fig5b}
\end{center}
\end{figure}

%For positive momentum $k$, $|\lambda_{+}|<|\lambda_{-}|$; while for negative $k$, we have $|\lambda_{+}|>|\lambda_{-}|$;
%
%Considering $|\lambda_{+}|>|\lambda_{-}|$ in small $D$ case, the minimum of $|\lambda_{+}|$ and maximum of $|\lambda_{-}|$ occur when
%\begin{align}
%{\cal E}^{'}_{3}= {AD} \cos{k}-2 \sqrt{1-D^2} \sin{k}=0
%%{\cal E}^{'}_{3}=\frac{ad}{b} \cos{k}-2 \sqrt{b^2-d^2} \sin{k}=0
%\end{align}
%with solution
%\begin{align}
%&\sin{k_c}=\frac{AD}{\sqrt{4  +A^2 D^2 -4 D^2}}\\
%%&\sin{k_c}=\frac{ad}{\sqrt{4 b^2 +a^2 d^2 -4 b^2 d^2}}\\
%&\cos{k_c}=\frac{2 \sqrt{1-D^2}}{\sqrt{4  +A^2 D^2 -4 D^2}}
%%&\cos{k_c}=\frac{2b\sqrt{b^2-d^2}}{\sqrt{4 b^2 +a^2 d^2 -4 b^2 d^2}}
%\end{align}
%If $|\lambda_{+}(k_c)|$ is smaller than unit, edge mode appears in momentum region,
%%If minimum of $|\lambda_{+}|$ is larger than unit, edge mode is absent.
%Therefore, critical condition for the presence or absence of edge mode reads
%\begin{align}
%|\lambda_{+}(k_c)|=1
%\end{align}
%For $|\lambda_{+}|<|\lambda_{-}|$, the corresponding critical
%condition reads $|\lambda_{-}(k_c)|=1$.
%

With further increase of $D$, 
we come close to the limiting case (ii), and have the reversed relation $|\lambda_{+}(k_c)|<|\lambda_{-}(k_c)|$. 
At $D=D_{c2}$, we obtain $|\lambda_{-}(k_c)|=1$.
Then the edge mode vanishes at $k=k_c$, but is still present on both sides of $k_c$.
Thus, with $D_{c2}<D<1$, edge mode contains two separate parts.
Fig.{\ref{fig5b}} shows the spectrum and $|\lambda_{\pm}|$ corresponding to this case.
The separate momentum regions, which are symmetric around $k_c$,
shrink simultaneously until $D=1$ is reached, where edge modes disappear completely.
In Fig.{\ref{fig5b}}  we obtain $k_c=1.307$ from Eq.(\ref{kc}).

\section{Zigzag edge}
\subsection{Analytic solution}
Let us now consider the zigzag edge geometry, as illustrated in
Fig.{\ref{fig1}}(b). Electrons are confined
in a diagonal strip: $1\le y-x \le N_{r}$, provided the edges are
placed at $y-x=1$ and $y-x=N_{r} $, normal to the
$(1,-1)$-direction. Translational invariance remains along the
$(1,1)$-direction. 
Bulk energy $E_{\rm b}$ in
terms of variables $\kappa = (k_x+k_y)/2$ and
$\xi = (k_x-k_y)/2$ is obtained from Eq.(\ref{bulk}) as
\begin{align}
E_{\rm b \pm}(\kappa,\xi) =& -4D+4D\cos{\kappa}\cos{\xi} \nonumber \\
&\pm \left[2A^2\left( \sin^2 \kappa \cos^2\xi + \cos^2 \kappa \sin ^2\xi
\right)\right.
\nonumber\\
 &+\left. \left(
 \Delta_B+4
 \cos \kappa \cos\xi  \right) ^2 \right]^{1/2}.
 \label{bulk11}
\end{align}
The Schr\"{o}dinger equation for the edge mode analogous to Eq.(\ref{TB2}) reads:
\begin{equation}
\hat{\cal E}_{11}\Phi_j +\hat t_{11} \Phi_{j-1}+\hat t^\dagger
_{11} \Phi_{j+1} = E_\uparrow (\kappa) \Phi_j, \label{TB11}
\end{equation}
where
\begin{align}
\hat{\cal E}_{11} &= -4D +\Delta_B \sigma_z,\\
\hat{t}_{11}
&=\frac{A}{\sqrt{2}}\sin {\kappa} \ \sigma_X -i \frac{A}{\sqrt{2}} \cos {\kappa} \ \sigma_Y, \nonumber\\
&+2 \cos
{\kappa} \ \sigma_z+2D \cos {\kappa},
\end{align}
and we have introduced 
\begin{align}
\sigma_{X}=\frac 1{\sqrt 2}\left( \sigma_x+\sigma_y \right)
\quad 
\sigma_{Y}=\frac 1{\sqrt 2}\left( \sigma_y-\sigma_x \right).
%\sigma_{Z}=\sigma_z
\end{align}
Here the conserved momentum is $\kappa=(k_x+k_y)/2$, with $-\pi/2<\kappa<\pi/2$.
We impose the boundary condition: $\Phi_0=\Phi_{N_r+1}=0$ for zigzag edge geometry,
and consider the thermodynamic limit $N_r\rightarrow\infty$. 

Assuming eigenstate of Eq.(\ref{TB11}) with property
$\Phi_j=\lambda\Phi_{j-1}=\lambda^{j}\Phi$, where $|\lambda|<1$,
%\cite{{Ed1},{Ed2}}, 
we obtain
\begin{align}
\left [\hat{\cal E}_{11} +\lambda \hat
t^\dagger_{11}(\kappa)+\lambda^{-1} \hat t _{11}(\kappa)\right]\Phi
& =P_{11}(\lambda,\kappa)\Phi= E_\uparrow (\kappa) \Phi \label{TB112}
\end{align}
In order to construct the annihilator $F_{11}$ for the zigzag edge, we
introduce further transformation: 
\begin{align}
\sigma_{\theta x}&=\cos{\theta} \ \sigma_X + \sin{\theta} \ \sigma_z, \\
\sigma_{\theta z}&=-\sin{\theta} \ \sigma_X +\cos{\theta} \ \sigma_z.
%\\\sigma_y&=\sigma_Y
\end{align}
Then, we have
\begin{align}
&\hat{\cal E}_{11} = -4D +\Delta_B \left( \sigma_{\theta x} \sin{\theta} + \sigma_{\theta z} \cos{\theta}\right),\\
&\hat{t}_{11}
=\left( \frac{A}{\sqrt 2} \sin{\kappa} \cos{\theta} +2   \cos{\kappa} \sin{\theta} \right) \sigma_{\theta x}-i\frac{A}{\sqrt 2} \sigma_y \cos{\kappa} 
\nonumber\\
& \hspace{2em} +\left( -\frac{A}{\sqrt 2} \sin{\kappa} \sin{\theta} + 2   \cos{\kappa} \cos{\theta} \right) \sigma_{\theta z} +2 D \cos{\kappa}.
\end{align}
%%%%%%%%%%%%%%
We decompose $P_{11}$ in Eq.(\ref{TB112}) 
as $P_{11}=H_{11}+F_{11}$ where
\begin{align}
H_{11}&=\left(\Delta_B \sin{\theta} \sigma_{\theta x}  
 -4 D\right)  
 \nonumber \\
 &+\left( \lambda+\lambda^{-1} \right) \left( \gamma_1 \ \sigma_{\theta x} +2 D \cos{\kappa} \right), \\
F_{11}&=\gamma_2 \ \sigma_y+\gamma_3 \ \sigma_{\theta z}, 
\label{F11}
\end{align}
with
\begin{align}
&\gamma_1=\frac{A}{\sqrt 2} \sin{\kappa} \cos{\theta} +2   \cos{\kappa} \sin{\theta},\\
&\gamma_2=-i\frac{A}{\sqrt 2} \cos{\kappa} \left(\lambda^{-1}-\lambda\right),\\
&\gamma_3=\Delta_B \cos{\theta} \nonumber \\
&+\left( \lambda + \lambda^{-1} \right) \left( -\frac{A}{\sqrt 2} \sin{\kappa} \sin{\theta} +2   \cos{\kappa} \cos{\theta} \right).
\end{align}
Note that $H_{11}$ contains non-Hermitian part due to the complex parameter $\lambda$. 
Since $H_{11}$
should have real eigenenergy $E_\uparrow (\kappa)$, 
we require the complex part in $H_{11}$ to be ineffective:
\begin{align}
\left(\gamma_1 \ \sigma_{\theta x} +2 D \cos{\kappa}\right)\Phi=0,
\end{align}
with $\sigma_{\theta x} \Phi=s \Phi$.  
Then the parameter $\theta$ is determined as
\begin{align}
%&\sin{\theta}=\frac{-s8D -\frac{\kappa}{|\kappa|} 
&\sin{\theta}=\frac{-8sD -{\rm sgn}\, \kappa
\sqrt{A^4 \tan^{4}\kappa+8 A^2 \left(1-D^2\right) \tan^{2}\kappa}}{8  +A^2 \tan^{2}{\kappa}}.
\label{sin}\\
&\cos{\theta}=\sqrt{1-\sin^2{\theta}} \ (>0).
\end{align}
The eigenenergy is given by
\begin{align}
E_\uparrow (\kappa)=s \Delta_B \sin{\theta}-4 D,
\end{align}
where $s=+1$ should be chosen for the edge mode, as shown below.
From Eq.(\ref{sin}) we obtain the limiting behavior:
\begin{align}
\sin\theta \rightarrow 
\begin{cases}
-{\rm sgn}\, \kappa & (\kappa\rightarrow \pm\pi/2) \\
-sD, & (\kappa\rightarrow 0)
\end{cases}
\end{align}
which gives 
$E_\uparrow (0)= -D\Delta$ and
$E_\uparrow (\pi/2)= -\Delta+4(1-D)$.

The corresponding annihilator $F_{11}$ in Eq.(\ref{F11}) should be proportional to 
$\sigma_y+ is \sigma_{\theta z}$, which requires
\begin{align}
\Delta_B \cos{\theta} +\left(\lambda+ \lambda^{-1}\right) R =s \frac{A}{\sqrt 2} \cos{\kappa} \left(\lambda^{-1}-\lambda\right).
\end{align}
with
\begin{align}
R=&-\frac{A}{\sqrt 2} \sin{\kappa} \sin{\theta} +2   \cos{\kappa} \cos{\theta}.
\label{R}
\end{align}
The solution is given by
\begin{align}
\lambda_{s\pm}(\kappa)=&\frac{-\Delta_B \cos{\theta}\pm\sqrt{\Delta^2_B \cos^2{\theta}+2 A^2 \cos^2{\kappa} -4 R^2}}{ 2R+ s\sqrt{2} A \cos{\kappa} },
\label{lambda11}
\end{align}
which leads,
for complex $\lambda$, to
\begin{align}
|\lambda_{s\pm}(\kappa)|^2=\frac{2R-s\sqrt{2} A \cos{\kappa}}{2R+s\sqrt{2} A \cos{\kappa}}.
\end{align}
In order to choose the appropriate branch out of $s=\pm 1$, we put $\kappa=0$ and obtain
$R=2\cos\theta >0$.
In this case the quantity
$$|\lambda_{s\pm}|^2=
(4  \cos{\theta}-s\sqrt{2} A)/(4  \cos{\theta}+s\sqrt{2} A)$$ 
is less than unity only with $s=+1$.  

We now consider the case $\kappa=\pm \pi/2$.
From Eq.(\ref{R}) we obtain $R=A/\sqrt{2}$, and then
\begin{align}
\lambda_{s\pm}(\pm \pi/2)=\pm i. \label{lb}
\end{align}
Since $\lambda_{s\pm}(\pm \pi/2)$ is complex, we obtain 
$|\lambda_{1\pm}|<1$ near the zone boundary.
This property means that edge mode near zone boundary is stable against transition from TI to SM (or OI).  
However, its existent region shrinks with increasing $D$. Hereafter, we use notation $\lambda_{\pm}=\lambda_{1\pm}(\kappa)$.

\subsection{Edge modes in TI and topologically trivial cases}

In this subsection
we fix $\Delta=\pm 1.2, A=1$,
which realizes TI with 
$0<D<1$ and $\Delta>0$. 
%Note that
%parameter $C$ is also trivial in zigzag edge case
%as in straight edge case. 
%We focus on the effect of particle-hole asymmetry by changing $D$.  
%With same sign of $B$ and $\Delta$, the zigzag edge system is  topological insulator and gapless point lies in $\Gamma$ point.
%In particle-hole symmetric case with $D=0$ \cite{{imura},{mao}}, edge mode spectrum is an odd function of $\kappa$, {\it i.e.} $E_\uparrow (\kappa)=-E(-\kappa)$ and parameter $\lambda$ is even function, {\it i.e.} $\lambda(\kappa)=\lambda(-\kappa)$ . Furthermore, there is reentrant edge mode near the boundary of 1D BZ when $\Delta<1.35$. 
%
Note that 
$\kappa=0$ is always a crossing 
point of two helical modes,
as a result of time-reversal symmetry of the system.
Hence this property is 
robust against particle-hole symmetry breaking as long as the system is still in TI.
However, the existent regions of edge mode will shrink with increasing $D$. 

\begin{figure}%[ht]
\begin{center}
\includegraphics[width=0.8\linewidth]{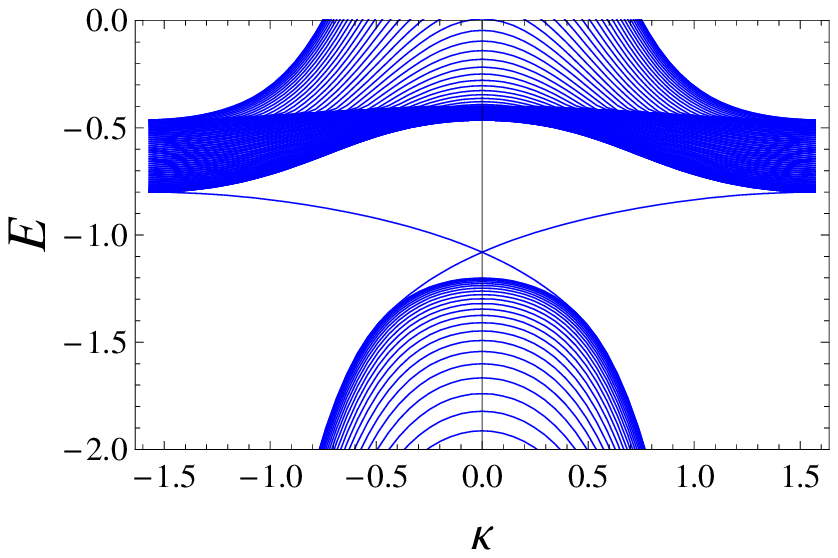}\\
\includegraphics[width=0.8\linewidth]{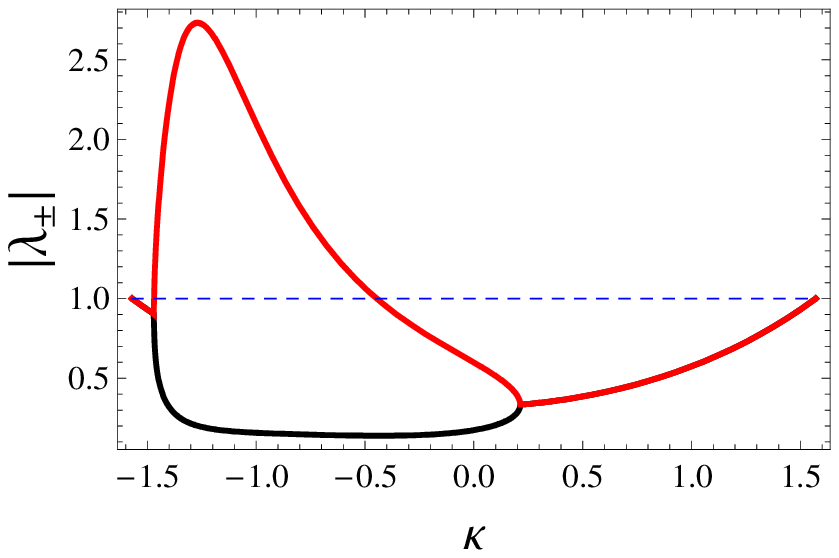}
\caption{ (Color online) Energy spectrum $E(\kappa)$ and parameters $|\lambda_{\pm}|$ in zigzag edge system with
$\Delta=1.2$, $A=1$,  $D=0.9$. 
%Here, $|\lambda_{\pm}|$ corresponds to right-going branch of edge mode.
} \label{fig7}
\end{center}
\end{figure}

Fig.\ref{fig7} shows the spectrum and $|\lambda_{\pm}|$ in the case of $D=0.9, \Delta=1.2$.
As shown in the lower panel,
we always have
$|\lambda_{\pm}|<1$
for $0<\kappa<\pi/2$, 
which means the presence of the edge mode
in the whole region of positive $\kappa$. 
With $\kappa<0$,
there appears a reentrant edge mode near the zone boundary $\kappa=-\pi/2$, 
which is not seen in the upper panel because of its much lower energy.
The binding energy of reentrant edge mode near zone boundary is expanded as
\begin{align}
E_{\uparrow}-E_{\rm b} \sim -4\Delta_{B}\left(\frac{D+1}{A}\right)^2 \left(\kappa+\frac{\pi}{2}\right)^2.
\label{binding}
\end{align}
The reentrant mode merges with bulk excitations at
momentum $\kappa_m$ that is given by the solution of 
\begin{align}
&(\Delta_B+4\cos\kappa)\cos\theta
-\sqrt{2} A\sin\kappa\sin\theta \nonumber\\
&= \Delta_B \cos\theta +2R 
=0.
\end{align}
The analytic expression of $\kappa_m$ in terms of system parameters is complicated and it's not illuminating.
For the parameters used in Fig.\ref{fig7}, we obtain the numerical value:
$\kappa_m = -0.936\pi/2$.
Since $\kappa_m$ is close to $-\pi/2$, the binding energy is only of the order of
$10^{-3}$ according to Eq.(\ref{binding}).  
The reentrant mode vanishes also at the zone boundary because of the relation: 
$|\lambda_{\pm}|\rightarrow 1$ as $\kappa\rightarrow -\pi/2$.

%%%%%%%%%%%%%%%%%%%%%%%%%%%%%%%%%%%%%%%%%%%%%%%%%%%%%%%%%%%%%%%
%%%%%%%%%%%%%%%%%%%%%%%%%%%%%%%%%%%%%%%%%%%%%%%%%%%%%%%%%%%%%%%%

\begin{figure}%[ht]
\begin{center}
\includegraphics[width=0.8\linewidth]{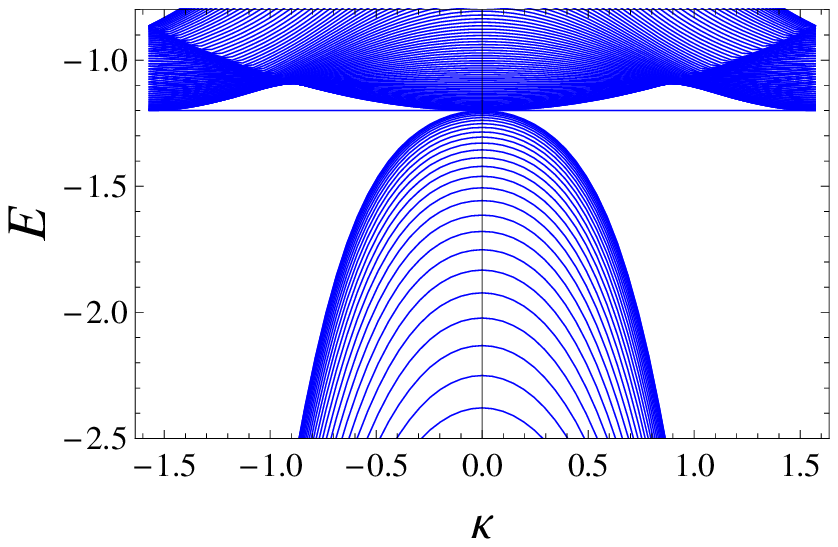}\\
\includegraphics[width=0.8\linewidth]{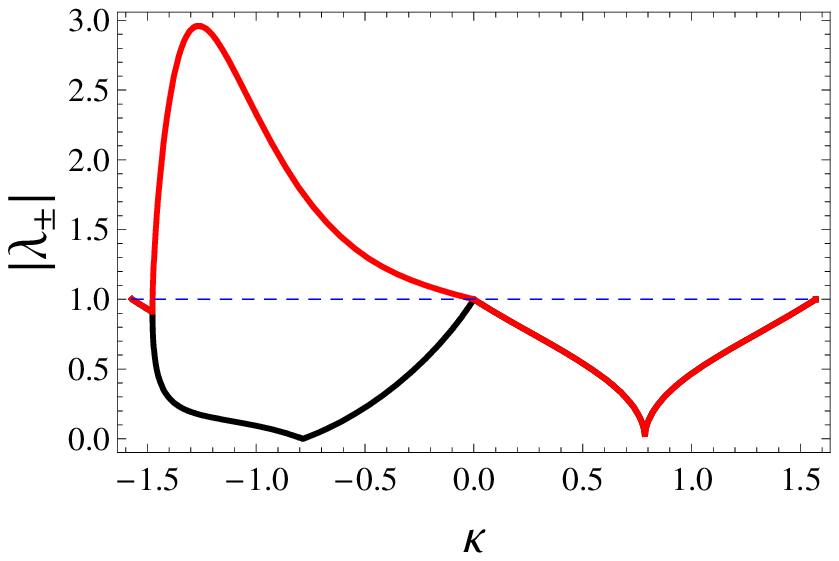}
\caption{ (Color online) 
The same quantities as in Fig.\ref{fig7} except for 
$D=1$ in this case.}
\label{fig8}
\end{center}
\end{figure}

In critical case with $D=1$, the bulk
energy gap closes, with extremum of $E_{\rm b}$ given by
\begin{align}
&E_{\rm b+}|_{min}=E_{\rm b+}(\pi/2, \pi/2)=-\Delta\\
&E_{\rm b-}|_{max}=E_{\rm b-}(0,0)=-\Delta
\end{align}
Fig.\ref{fig8} shows the spectrum and parameters $|\lambda_{\pm}|$ in this case.
Since we obtain $|\lambda_{\pm}(\kappa=0)|=1$ with  $D=1$, 
the crossing point of two helical edge modes merges with bulk excitations
associated with indirect gap-closing, 
However, edge mode still exists in the other momentum region with $|\lambda_{\pm}|<1$. 
Note that the right-going helical mode has the flat spectrum for $\kappa>0$,  and
that the left-going helical mode has the flat spectrum for $\kappa < 0$, 
%%%%%%%%%%%%%%%%%%%%%%%%%%%%%%%%%%%%%%%%%%%%%%%%%%%%%%%%%%%%%%%
%%%%%%%%%%%%%%%%%%%%%%%%%%%%%%%%%%%%%%%%%%%%%%%%%%%%%%%%%%%%%%%%

\begin{figure}%[ht]
\begin{center}
\includegraphics[width=0.8\linewidth]{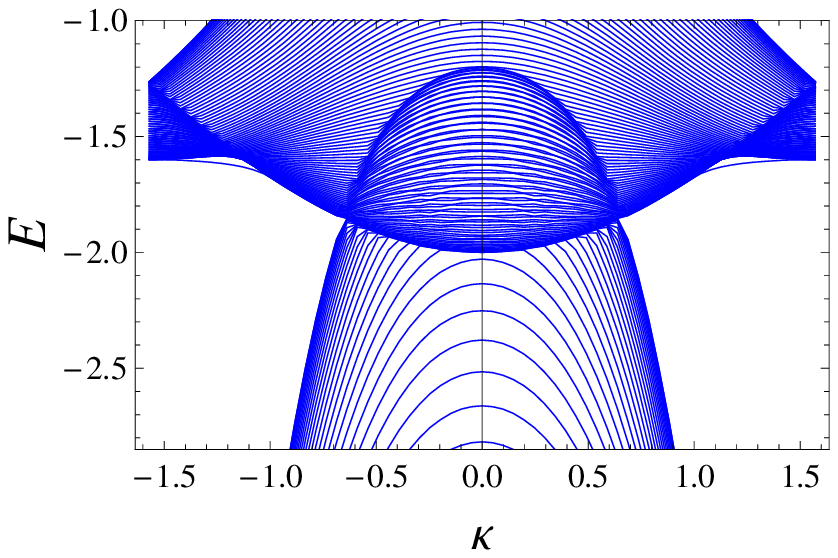}\\
\includegraphics[width=0.8\linewidth]{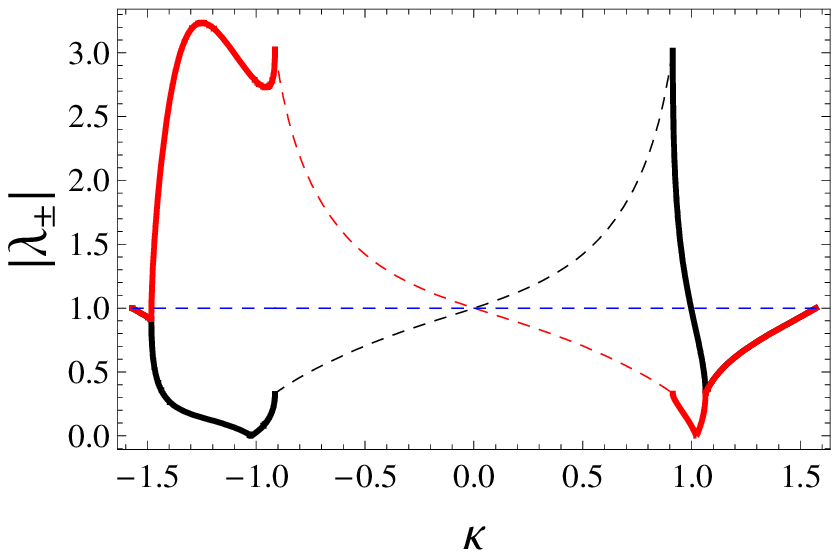}
\caption{ (Color online) 
The same quantities as in Figs.\ref{fig7} and \ref{fig8} except for 
$D=1.1$ in this case.
The dashed lines in the lower panel correspond to complex $\sin\theta$ which is unphysical.
}

\label{fig9}
\end{center}
\end{figure}

With $D>1$,  the system is no longer a
topological insulator.
However the edge mode still survives in some
momentum region. 
Fig.\ref{fig9} shows the 
spectrum and parameters $|\lambda_{\pm}|$ in this case.
The edge mode contains two separate parts lying near
the zone boundary in valence and conduction bands. 
This new phase 
%may be called helical metal; it 
is bulk metal but with helical edge mode. 
Note that the solution 
$\sin {\theta}$ becomes complex in certain momentum region,
which is unphysical.
%Then $|\lambda_{\pm}|$ in this region is not physical.

%%%%%%%%%%%%%%%%%%%%%%%%%%%%%%%%%%%%%%%%%%%%%%%%%%%%%%%%%%%%%%%
%%%%%%%%%%%%%%%%%%%%%%%%%%%%%%%%%%%%%%%%%%%%%%%%%%%%%%%%%%%%%%%%

\begin{figure}%[ht]
\begin{center}
\includegraphics[width=0.8\linewidth]{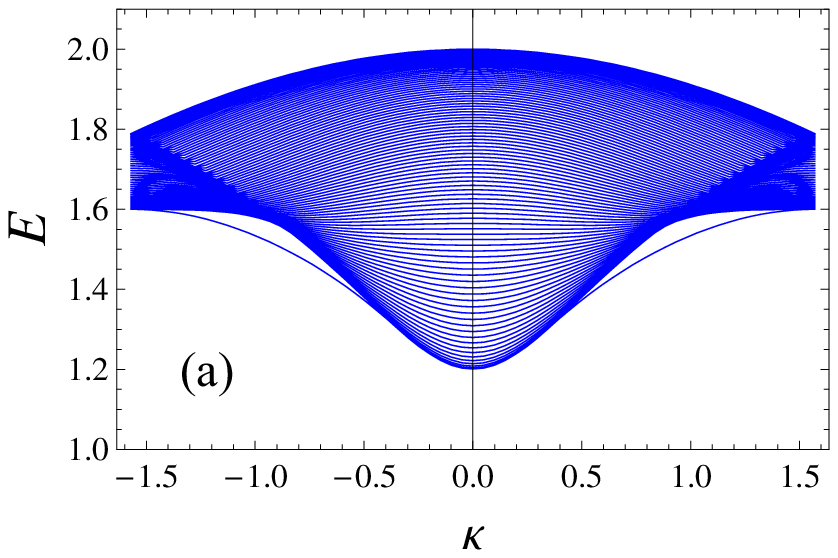}\\
\includegraphics[width=0.8\linewidth]{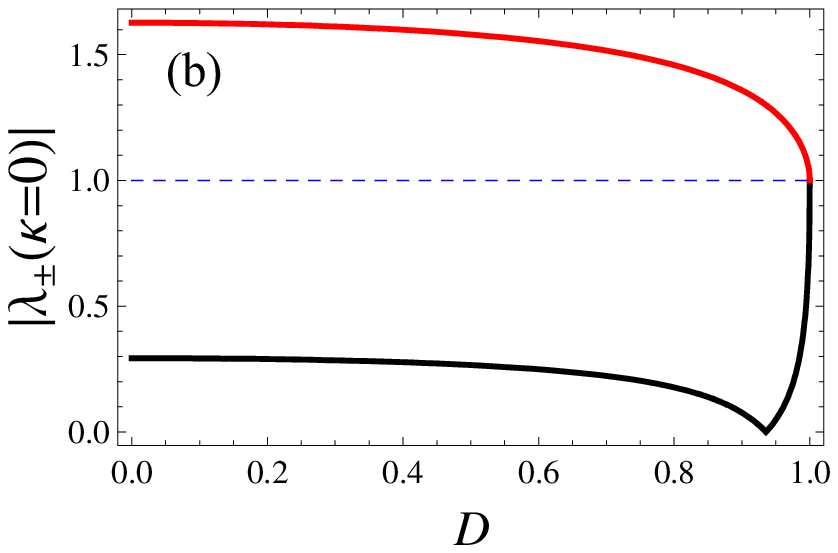}
\caption{ (Color online) (a) Conduction band $E (\kappa)$ with $\Delta=-1.2$, $A=1$,  $D=0.9$
in zigzag edge system. (b) Parameter $|\lambda_{\pm}(\kappa=0)|$ as a function of $D$.} 
\label{fig10}
\end{center}
\end{figure}

%\subsection{Edge mode in ordinary insulator phase}
Let us now consider the OI case with $\Delta=-1.2$.
According to Eq.(\ref{lb}), edge mode will appear near the
boundary of BZ. 
Fig.{\ref{fig10}} (a) shows an example of the spectrum (upper panel).
There is no edge mode within the energy gap.
Fig.{\ref{fig10}} (b) shows the absence of the edge mode at $\kappa=0$,
since
at least one of the
$|\lambda_{\pm}|$ is larger than unity.
%%%%%%%%%%%%%%%%%%%%%%%%%%%%%%%%%%%%%%%%%%%%%%%%%%%%%%%%%%%%%%%
%%%%%%%%%%%%%%%%%%%%%%%%%%%%%%%%%%%%%%%%%%%%%%%%%%%%%%%%%%%%%%%%

\section{Summary}
In this paper, we have analyzed edge modes in the particle-hole asymmetric BHZ model taking both straight and zigzag edges.
Our particular attention has been the survival or disappearance
of helical edge modes through transition from TI to SM, and also its presence in OI.

For straight edge, 
with increasing particle-hole asymmetry controlled by $D$, 
the edge mode 
breaks up into two parts in momentum space, and each part shrinks simultaneously.
Associated with transition from TI to SM, the system becomes
normal SM without edge mode. 
On the other hand, in zigzag edge case, helical
edge mode is robust against particle-hole symmetry breaking. 
In TI, gapless point always lies at the $\Gamma$ point and 
reentrant edge mode appears near the boundary of 1D BZ. 
With closing of indirect gap in bulk energy band, 
those edge modes
near the crossing point of Kramers'
pair disappear, while the edge mode near the zone boundary
remains. 

%The sign change of $\Delta$ leads the system into OI.
In OI systems, helical edge
mode, if exists, should not be inside the energy gap.
Then such edge mode does not change
the topology of energy band from OI without edge modes. 
In straight edge
system, the presence of edge mode is sensitive to parameters. However,
for zigzag edge case, it is insensitive to values of parameter because 
the edge mode always merges with bulk modes
at the zone boundary.

In summary, we have demonstrated the presence of
helical edge modes outside the energy gap
as a novel state in particle-hole asymmetric BHZ model. 
%It confronted us with questions about its physical origin and experimental realization. 
Our analytic approach to study the helical modes
%which is the character of TI, with the bulk SM or OI state,
can be generalized to 3D systems.
The results will be reported elsewhere. 
%we shall have more intuition on novel properties of actual systems.

\bibliography{apssamp}% Produces the bibliography via BibTeX.

\begin{thebibliography}{99}

\bibitem{he2}
C. Wu, B.A. Bernevig, and S.C. Zhang: Phys. Rev. Lett. 96, 106401
(2006).

\bibitem{he3}
C. Xu and J. Moore: Phys. Rev. B 73, 045322 (2006).

\bibitem{ex1}
D. Hsieh, Y. Xia, L. Wray, D. Qian, A. Pal, J.H. Dil, J. Osterwalder, F. Meier, G. Bihlmayer, C.L. Kane, Y.S. Hor, R.J. Cava, and M.Z. Hasan: Science 323, 919 (2009).

\bibitem{ex2}
D. Hsieh, L.A. Wray, D. Qian, Y. Xia, Y.S. Hor, R.J. Cava, and M.Z. Hasan: arXiv:1001.1574.

\bibitem{ex3}
A.A. Taskin and Y. Ando: Phys. Rev. B 80, 085303 (2009).

\bibitem{ex4}
A.A. Taskin, K. Segawa, and Y. Ando:  Phys. Rev. B 82, 121302 (2010).

\bibitem{ex5}
Y.S. Hor, P. Roushan, et al.:  Phys. Rev. B 81, 195203 (2010).

\bibitem{ex6}
J.G. Analytix, J.H. Chu, Y. Chen, F. Corredor, R.D. McDonald, Z.X. Shen, and I.R. Fisher:  Phys. Rev. B 81, 205407 (2010).

\bibitem{ex7}
K. Eto, Z. Ren, A.A. Taskin, K. Segawa, and Y. Ando:  Phys. Rev. B 81, 195309 (2010).

%\bibitem{hm}
%D.L. Bergman and G. Refael: arXiv:1003.3018.
\bibitem{zhang4}
H. Zhang, C.X. Liu, X.L. Qi, X. Dai, Z. Fang and S.C. Zhang:
Nat. Phys. 5, 438 (2009).

\bibitem{BHZ}
B. A. Bernevig, T. L. Hughes and S.C. Zhang: Science 314, 1757
(2006).

\bibitem{cm}
B. Zhou, H.Z. Lu, R.L. Chu, S.Q. Shen, and Q. Niu: Phys. Rev. Lett. 101, 246807 (2008).

\bibitem{cm2}
E.B. Sonin:  Phys. Rev. B 82, 113307 (2010).

\bibitem{jpsj}
M. K\"{o}nig, H. Buhmann, L.W. Molenkamp, T.L. Hughes, C.X Liu, X.L
Qi and S.C Zhang: J. Phys. Soc. Jpn \textbf{77}, 031007 (2008).

\bibitem{imura}
K. Imura, A. Yamakage, S.J. Mao, A. Hotta and Y. Kuramoto: Phys. Rev. B 82, 085118 (2010).

\bibitem{mao}
S.J. Mao, Y. Kuramoto K. Imura, and A. Yamakage:
arXiv:1008.0481; J. Phys. Soc. Jpn in press.


\bibitem{gapless}
S. Murakami: arXiv:1006.1188.

\bibitem{Ed1}
M. Creutz and I. Horvath:Phys. Rev. D 50, 2297 (1994).

\bibitem{Ed2}
M. Creutz: Rev. Mod. Phys. 73, 119 (2001).




\end{thebibliography}

\end{document}